\documentclass[runningheads]{llncs}

\usepackage{authblk}

\usepackage{graphicx}

\usepackage{comment}
\usepackage{xcolor}
\usepackage{amsmath}
\usepackage{amssymb}

\usepackage{subcaption}

\usepackage{booktabs}
\usepackage{tabularx}
\usepackage{svg}
\usepackage{dependencies}
\usepackage{float}

\usepackage[hidelinks]{hyperref}

\newcommand{\repthanks}[1]{\textsuperscript{\ref{#1}}}
\makeatletter
\patchcmd{\maketitle}
  {\def\thanks}
  {\let\repthanks\repthanksunskip\def\thanks}
  {}{}
\patchcmd{\@maketitle}
  {\def\thanks}
  {\let\repthanks\@gobble\def\thanks}
  {}{}
\newcommand\repthanksunskip[1]{\unskip{}}
\makeatother

\begin{document}

\title{Liquid Staking Tokens in \\Automated Market Makers}

\author{Krzysztof Gogol\thanks{Both Authors contributed equally to this work\protect\label{X}} \inst{1,4} \and
Robin Fritsch\repthanks{X} \inst{2} \and
Malte Schlosser\inst{1}
\and
Johnnatan Messias\inst{4}
\and
Benjamin Kraner\inst{1}
\and
Claudio Tessone\inst{1}\inst{3}
}

\authorrunning{K. Gogol, R. Fritsch et al.}
\institute{University of Zurich\and
ETH Zurich \and
UZH Blockchain Center \and
Matter Labs}

\maketitle              %
\begin{abstract}
This paper studies liquid staking tokens (LSTs) on automated market makers (AMMs), both theoretically and empirically.
LSTs are tokenized representations of staked assets on proof-of-stake blockchains. 

First, we model LST-liquidity on AMMs theoretically, categorizing suitable AMM types for LST liquidity and deriving formulas for the necessary returns from trading fees to adequately compensate liquidity providers under the particular price trajectories of LSTs.
For the latter, two relevant metrics are considered: (1) losses compared to holding the liquidity outside the AMM (loss-versus-holding, or “impermanent loss”), and (2) the relative profitability compared to fully staking the capital (loss-versus-staking) which is specifically tailored to the case of LST-liquidity.

Next, we empirically measure these metrics for Ethereum LSTs across the most relevant AMM pools.
We find that, while trading fees often compensate for impermanent loss, fully staking is more profitable for many pools, raising questions about the sustainability of the current LST liquidity allocation to AMMs.

\end{abstract}

\section{Introduction}

Liquid Staking Tokens (LSTs) have become the preferred method of staking, now constituting over a third of total staked ETH  \cite{2023DuneETHStaking}.
Moreover, liquid staking has grown to become the largest category in decentralized finance (DeFi) by allocated capital. 
DeFi itself has experienced remarkable growth in recent years with over 5 million private wallet addresses interacting with DeFi protocols \cite{2022StatistaUsers}, and investment and central banks are experimenting with its application in traditional finance \cite{BankforInternationalSettlements.2023TheRisks,BankforInternationalSettlements.2023LessonsGovernors}. While the emergence of liquid staking has been one of the DeFi ecosystem's most significant recent developments, it has so far seen limited academic attention.

In traditional staking on proof-of-stake (PoS) blockchains \cite{saleh2020pos}, stakers lock (``stake'') native blockchain tokens with a validator, typically subject to a lock-up period. Validators maintain the blockchain network and receive staking rewards in exchange. Stakers have the option to run their own validators or become customers of a staking pool operator. In the case of the Ethereum blockchain, the minimum collateral of 32 Ether (ETH) is required to set up and operate a validator.

In contrast, liquid staking tokens (LSTs) allow for more simple and flexible staking. LSTs are tokenized representations of the staked underlying asset, and can be minted and redeemed for the corresponding amount of the underlying asset directly from the LST protocol.\footnote{Some blockchains, such as Cardano, come with ``native liquid staking'' without the need for LST protocols.}
Additionally, since LSTs are fungible tokens, they can be bought and sold on exchanges, in particular on decentralized exchanges running automated market maker protocols (AMMs).

LST protocols manage a network of validators and take a percentage of the staking rewards as compensation for their services \cite{Gogol2024SoK:LST}.
Holders of LSTs can earn staking rewards without directly locking any tokens in their own validator or with an external validator operator. Moreover, LSTs enable combining the benefits of staking with other functionalities within DeFi. 

In particular, the emergence of LSTs has transformed decentralized lending, with LSTs becoming the dominant collateral \cite{2023MessariQ123Report}, and has also reshaped investment strategies within DeFi.
A unique advantage and feature of LSTs which is vital to their intended functionality is the possibility to trade them on AMMs, which enables instant entering or exiting of staking positions.
Precisely this aspect, i.e.\ LST liquidity on AMMs, is the subject of this research.

We systematically study -- both theoretically and empirically -- the functioning of LST-liquidity on AMMs.
A central aspect in this regard is the profitability of providing LSTs as liquidity to AMMs.

On AMMs, LSTs are commonly traded against their underlying asset.
Since the price of LSTs is by design either pegged to a unit of the underlying asset or experiences a constant increase compared to the underlying asset as a result of staking rewards, such LST trading pairs exhibit a specific behavior: Apart from short-term price fluctuations, their price changes are constant and predictable.
This paper studies AMM liquidity in this particular setting.

\subsection{Methodology and Contribution}

We begin by categorizing suitable AMM types for LST liquidity.
Subsequently, we concentrate on the profitability of liquidity providers (LPs).
We theoretically derive formulas for the necessary returns from trading fees as a function of the staking rate.
For this purpose, we consider the following two metrics measuring the losses to LPs on AMMs.

First, we consider a well-established metric for LP profitability on AMMs: loss-vs-holding (LVH) -- also known as ``impermanent loss'' or ``divergence loss''.
It measures how much a liquidity position loses from changing prices compared to keeping the initial position outside the AMM.

A second relevant measure, that we introduce specifically for LSTs, is comparing providing LST liquidity on an AMM to fully staking the capital.
We coin the term \emph{loss-versus-staking} (LVS) for this metric.
Instead of providing an LST and its underlying assets as liquidity on an AMM, LPs can stake their capital by fully holding it in LSTs. In the latter case, LPs earn staking rewards on their full capital compared to trading fees from the AMM and staking rewards on part of their capital in the first case.

For each metric, we derive the formula for the necessary returns from trading fees required to adequately compensate LPs for each of the losses under the particular price trajectory of LSTs.
This results in expressions for the required returns as a function of the staking rate $r$.

Subsequently, we empirically measure these metrics for the largest Ethereum LSTs: stETH and wstETH by Lido \cite{2020Lido:Whitepaper}, Coinbase's cbETH \cite{2022CoinBaseWhitepaper} and Rocketpool's rETH \cite{2023RocketPool}.
Together, they represent over 80\% of TVL in liquid staking\footnote{\url{https://defillama.com/lsd}}.
For these tokens, we consider the largest AMM pools in terms of TVL, which are on the decentralized exchanges Curve (v1 \cite{Egorov2019StableSwap-efficientLiquidity} and v2 \cite{Egorov2021AutomaticPeg}) and Uniswap (v3 \cite{Adams2021UniswapCore}) on the Ethereum blockchain \cite{Buterin2014Ethereum:Platform.}.

We find that simply holding LSTs is more profitable than allocating LSTs to most AMM liquidity pools: trading fees on the AMMs did not compensate LPs for not staking their capital fully (by allocating it to LSTs) during the studied time period. Yet, impermanent loss was sufficiently compensated for most of the analyzed liquidity pools. Nonetheless, our result put the sustainability of the current allocation of LST liquidity on AMMs into question. 
This is particularly noteworthy since the amount of AMM liquidity is already relatively low compared to the overall supply of LSTs at the time of writing (\$0.5 billion liquidity compared to more than \$30 billion supply, cf.\ Figure \ref{fig:TVLVolume}).

\section{Related Work}
Liquid staking has received limited attention in the literature so far. The price difference between  LSTs and their underlying cryptocurrency has been investigated by Scharnowski et al.\ \cite{Scharnowski2022LiquidDiscovery}. A taxonomy for LSTs, including the differentiation between rebase, reward and dual token models, can be found in the SoK \cite{Gogol2024SoK:LST}
LSTs' tracking of staking rewards has been empirically examined in~\cite{Gogol2023EmpiricalProtocols}. 
Furthermore, leveraging LSTs on lending protocols and the principal-agent problem in liquid staking have been discussed in \cite{xiong2023leverageLST} and \cite{Tzinas2023AgentLST}, respectively.

Liquidity provider returns have been discussed in the literature since the first theoretical analysis of the current form of AMMs \cite{Angeris2021Analysis}.
A comprehensive theoretical treatment of slippage and impermanent loss in different types of AMMs, including Uniswap v3 and Curve v1 which are of interest for this paper, can be found in the SoK by Xu et al.~\cite{Xu2021SoK:Protocols}.
Furthermore, a general framework for impermanent loss has been laid out in \cite{Tiruviluamala2022AMakers}.  

On the empirical side, the general profitability of liquidity providers on AMMs has been studied in the literature for different AMMs, token pairs, and time periods: for Uniswap v2 in \cite{Heimbach2021BehaviorExchanges} and for Uniswap v3 in \cite{Fritsch2021ConcentratedMakers,Heimbach2022RisksProviders,Loesch2021ImpermanentV3}. Most of these works focus on impermanent loss (loss-versus-holding) as a measure of LP profitability.
As an alternative metric, loss-versus-rebalancing (LVR) that benchmarks AMM liquidity against a rebalancing portfolio that hedges the position's market risk was introduced in \cite{Milionis2022AutomatedLoss-Versus-Rebalancing,milionis2023automated}, and used to measure LP profitability empirically in \cite{fritsch2024profitability}.

\section{Background}

\subsection {Liquid Staking Tokens}
LSTs are tokenized representations of staked native PoS blockchain tokens. They can be redeemed for the underlying token at any time (possibly with a delay) at the LST protocol, or traded on AMMs. Consequently, their market price on AMMs oscillates around the value of the staked assets they represent. Like other pegged tokens such as stablecoins (most commonly pegged to fiat currency) and wrapped tokens (pegged to tokens from other blockchains), LSTs offer a means to represent staked tokens in a liquid and tradable form. LSTs can be categorized into two main types based on their staking distribution models: rebase or reward \cite{Gogol2023EmpiricalProtocols}. The third model - dual token LSTs - is disappearing \cite{2023StakeWise}. LSTs also vary in validator selection processes and governance \cite{Gogol2024SoK:LST}. Additionally, liquid staking protocols typically charge a fee, usually around 10-25\% of the staking rewards, to cover operational costs and provide incentives for the platform's maintenance and development.

\subsubsection{Rebase-LST.} In the rebase model, LST holders receive staking rewards in the form of additional LSTs on a daily basis. Meanwhile, the rebase-LST tokens always remain redeemable for 1 ETH (or 1 unit of its underlying asset), meaning its price stays close to 1 ETH.
A prominent example is stETH from the Lido Protocol \cite{2020Lido:Whitepaper}, which was the first LST and currently holds the largest market capitalization among liquid staking tokens. The historical values of rebase-LST, denominated in ETH, are presented in Figure \ref{fig:RebaseLST}.

The rebase model is not compatible with certain DeFi protocols such as lending pools like Aave \cite{2020AaveV1.0} and Compound \cite{RobertLeshner2019Compound:Protocol}. These platforms do not support stETH since lenders to stETH pools would only earn staking rewards from stETH tokens in the pool that have not been borrowed. Moreover, rebase-LSTs are not compatible with some AMMs, Uniswap being the most notable example \cite{2023UniswapIntegration,2023LidoRebase}.

\subsubsection{Reward-LST.} For reward-LSTs, the staking rewards are accumulated in the value of the staking tokens. That is, one reward-LST token is always redeemable for its initial underlying amount (e.g.\ 1 ETH) as well as all staking rewards earned up to this time. Hence, the value of reward-LSTs measured in ETH will increase over time.
Examples of reward-LSTs include wstETH from Lido\cite{2020Lido:Whitepaper}, rETH from RocketPool \cite{2023RocketPool} and cbETH from CoinBase \cite{2022CoinBaseWhitepaper}. The historical prices of reward-LSTs are presented in Figure \ref{fig:RewardsLST} and compared to the returns from staking ETH.

\subsection{Risks of LSTs}
LSTs carry a risk of slashing, associated with the staking process. However, this risk is mitigated to some extent since the total staked tokens are distributed across multiple validators, making it less pronounced compared to traditional staking. Notably, certain liquid staking protocols like RocketPool mandate validators to deposit collateral, which is then used to cover losses from slashing\cite{2023RocketPool}.

The market prices of LSTs exhibit fluctuations around the peg price (derived from the reserves of staked tokens)\cite{Gogol2023EmpiricalProtocols}. As with any pegged token, there exists a de-peg risk, \eg in the aftermath of significant market events.
The effects of the Terra/Luna crash and the FTX insolvency in 2022 on LST prices can be seen in Figure \ref{fig:RebaseLST} for rebase-LSTs, and Figure \ref{fig:RewardsLST} for reward-LSTs.

In general, redemptions of LSTs for the underlying assets are limited by the unstaking process of the PoS blockchain, which can include waiting periods and withdrawal limits per time period. Hence, large demand for unstaking can lead to delays in LST redemptions through the LST protocol, and therefore cause LST prices to depeg.

\subsection{Constant Function Market Makers (CFMMs)}
AMMs are protocols deployed on a blockchain to allow traders to exchange assets in a fully noncustodial manner.
Their main component are liquidity pools for a number (most commonly two) of traded tokens, into which liquidity providers (LPs) deposit reserves.
Traders can then swap these assets against the pool at an exchange rate determined by an AMM-specific trade function, depending on the reserves currently in the pool. For every trade, a trading fee in form of a fixed percentage of the trade is paid. These fees (or, for some AMMs, part of them) are distributed pro rata among liquidity providers in the pool.

Constant Function Market Makers (CFMM), the most common form of AMMs, have a trading function $\varphi: \mathbb{R}^N_+ \rightarrow \mathbb{R}$, often called a conservation function or reserve curve, associated to them, which maps their token reserves to a real number, often referred to as invariant, which is kept constant for each trade. Here $N$ denotes the number of tokens in the liquidity pool, which for all relevant LST pools (and almost all relevant AMM pools in general) is $N=2$.

A wide range of different CFMMs has been implemented and suggested \cite{Xu2021SoK:Protocols}. In the following, we describe the major types of CFMMs, which also include those being used for LST trading, at the time of writing.
For all, the pools reserves are denotes by $x_1,\ldots,x_N$.

    \subsubsection{Constant Product (CPMM).}
        The first-used and most simple invariant, used in Uniswap v1 and v2 \cite{Adams2020UniswapCore}:
            $$\prod_{i=1}^N x_i = L^N.$$
    \subsubsection{Constant Product with Concentrated Liquidity (CLMM).}
        Introduced by Uniswap v3 \cite{Adams2021UniswapCore}), it allows LPs to specify a price range $[p_a,p_b]$ to which they provide liquidity. Then the trade invariant in this interval is
            $$\left( x_1+\frac{L}{\sqrt{p_b}} \right)\left(x_2+L\cdot \sqrt{p_a}\right) = L^2.$$
    \subsubsection{Stableswap Invariant.}
        Introduced with Curve v1 \cite{Egorov2019StableSwap-efficientLiquidity}, this CFMM is specifically designed for assets that trade at a constant price to each other. The invariant is
        \begin{equation}\label{eq:curvev1}
            K\cdot D^{N-1}\cdot \sum\limits_{i=1}^N x_i +\prod\limits_{i=1}^N x_i = K\cdot D^N + \left(\frac{D}{N} \right)^N
        \end{equation}
        with
            $$K= \frac{A\cdot\prod_{i=1}x_i}{D^N}\cdot N^N,$$
        where $A$ is a concentration parameter.
    \subsubsection{Cryptoswap Invariant.}
        Curve v2 \cite{Egorov2021AutomaticPeg}, an adaptation of the Stableswap invariant for assets with non-constant prices. It uses the same equation as (\ref{eq:curvev1}), but with a different $K$, defined as
            $$K = A\cdot \underset{\text{ =: }K_0}{\underbrace{\frac{A\cdot\prod_{i=1}x_i}{D^N}\cdot N^N}} \cdot \frac{\gamma^2 }{(\gamma+1- K_0)^2}.$$

\section{The Theory of LSTs on AMMs}

While LSTs can potentially be paired and traded against any other token on AMMs, the majority of LST liquidity lies in liquidity pools containing the LST and the underlying token.
This supports the main purpose of LSTs on AMMs: Entering or exiting staking positions by buying or selling the LST for the underlying.
Pairing LSTs against their underlying is also advantageous, since the volatility between the two should be low (if the LST functions well).

\subsection{Suitable types of AMMs for LSTs}

Different types of AMMs come with different properties and limitations, rendering them more or less well suited for specific liquid staking token pairs.
Refer to Table \ref{tab:LiquidStakingInAMM} for which AMMs are most suitable for rebase and reward LSTs to achieve high capital efficiency for LPs and low price impact for traders.
While LSTs could potentially be traded on CPMMs, their low capital efficiency for assets that move little in price, such as LSTs, makes them an unattractive choice.

Rebase-LSTs maintain a price pegged to 1 ETH, aligning perfectly with the Stableswap invariant, which was specifically designed for pairs with stable prices. 
Although CLMMs are theoretically also well-suited for trading pairs with stable prices, they are currently not widely used for trading rebase-LSTs due to the technical incompatibility of Uniswap v3 -- the market-leading CLMM - with rebase-LSTs \cite{2023UniswapIntegration,2023LidoRebase}.

Reward-LSTs, on the other hand, increase their value every day, according to the staking rate. Consequently, they are ideally traded on AMMs employing the Cryptoswap invariant or a CLMM. It is important to note that with CLMMs, periodic rebalancing of liquidity positions is necessary.

\begin{table}[!htb]
\centering
\begin{tabularx}{\linewidth}{XXX} 
    \textbf{Token 1} & \textbf{Token 2} & \textbf{AMMs} \\
    \toprule 
     rebase-LST & ETH & Stableswap \\
    \hline
   rebase-LST & rebase-LST & Stableswap \\
    \hline
     reward-LST & ETH & Cryptoswap, CLMM(*) \\
    \hline
     reward-LST & reward-LST & Cryptoswap, CLMM \\
    \bottomrule
\end{tabularx}
\caption{Suitable (in terms of liquidity efficiency) AMMs for trading pairs with different LST types. (*) regular rebalancing required}
\label{tab:LiquidStakingInAMM}
\end{table}

\subsection{Losses and Required Returns for LPs}

Within any AMM pool, it is imperative that trading fees adequately reward liquidity providers for their capital contribution.
On the one hand, LPs earn trading fees, on the other hand, the value of their liquidity position fluctuates with token prices. In particular, when prices change, LPs suffer a loss when comparing the value of their liquidity position to holding their liquidity elsewhere than in the AMM pool.

At the very least, returns from trading fees should offer sufficient compensation to offset potential losses of a liquidity position compared to a number of basic benchmark strategies. In this context, a generally relevant metric is the loss versus holding the initial positions outside the AMM (loss-vs-holding, LVH, sometimes referred to as ``impermanent loss'')
Specifically for AMM pools involving liquid staking tokens and their underlying assets, it is furthermore highly relevant to measure the loss compared to a fully staked portfolio (loss-vs-staking, LVS).

Each of these losses can be quantified specifically for the particular price trajectory of LST trading pairs. This allows us to calculate the required returns to compensate LPs as a function of the staking rate $r$.
Moreover, in broad terms, a liquidity provider's return from trading fees can be expressed as follows:
\begin{equation*}
    r_{LP} =\frac{\text{trading volume}\cdot \text{trading fee}}{\text{liquidity in pool}}
\end{equation*}
This implies that, based on the required returns, the necessary trading volumes (relative to the pool size) to render the pools profitable for LPs can be inferred.

It is important to note that for CLMMs, LPs earn different returns depending on the concentration of their positions and whether their liquidity is consistently within range. Consequently, one must factor in the concentration factor of the position and consider only the amount of liquidity in range when determining the denominator in the equation above.\\

In the following, we theoretically derive closed-form expressions for the different LP losses -- and thereby the required returns to compensate them -- for the particular case of LST liquidity as a function of the staking rate $r$. 

We consider a liquidity pool containing an LST and the underlying asset (such as ETH), and use the underlying asset as unit of measurement. Let $P(t)$ denote the price of the LST (denominated in the underlying asset) at time $t\geq 0$.
As a baseline, we consider an AMM liquidity position without taking earnings from trading fees into account.
Let the liquidity position consist of $x(t)$ LST units and $y(t)$ units of the underlying token at time $t$.
Then its value is given by
\begin{align*}
    V_{LP}(t) = x(t) P(t) + y(t).
\end{align*}
Furthermore, let $V_{HOLD}(t)$ denote the value of holding the initial position:
\begin{align*}
    V_{HOLD}(t) = x(0) P(t) + y(0)
\end{align*}
Next, we consider the value $V_{LST}(t)$ of a fully staked position (i.e.\ holding the initial position fully in the LST):
\begin{align*}
    V_{LST}(t) = V_{LP}(0) \frac{P(t)}{P(0)} = \left( x(0) + \frac{y(0)}{P(0)} \right)P(t)
\end{align*}
Comparing the AMM position value to these benchmarks (holding and staking) reveals the size of the losses for LPs:
\begin{align*}
    LVH(t) &=  V_{HOLD}(t) - V_{LP}(t) \\
    LVS(t) &=  V_{LST}(t) -  V_{LP}(t)
\end{align*}
Moreover, the required returns from trading fees $rr_{LVH}$ and $rr_{LVS}$ needed to compensate for LVH and LVS, are given as follows:
\begin{align*}
    rr_{LVH}(t) &= \frac{V_{HOLD}(t)}{V_{LP}(t)} - 1 \\
    rr_{LVS}(t) &= \frac{V_{LST}(t)}{V_{LP}(t)} - 1
\end{align*}

\subsection{Required returns for rebase LSTs}

For trading pairs between rebase LSTs and the underlying asset, the matter is relatively straightforward, as the price of rebase LSTs is pegged to the price of the underlying asset.
We only briefly discuss this case at a high level.

First, note that in most AMMs half of the initial position is allocated to the LST and will be earning rewards when holding. Consequently, if the value of the initial position is $V_{LP}(0)$, the value of holding becomes $V_{HOLD}(t) = V_{LP}(0)(1/2 + e^{rt}/2)$.

On the other hand, due to the stable price, the pool maintains an equal allocation between LST and the underlying. Therefore, half of the liquidity in the pool is earning rewards, resulting in $V_{LP}(t) = V_{LP}(0)(1/2 + e^{rt}/2)$. As a result, no loss-versus-holding occurs, and $rr_{LVH}(t)=0$.

Finally, the value of the fully staked portfolio is $V_{LST}(t)=V_{LP}(0)e^{rt}$, implying
\begin{equation*}
    rr_{LVS}(t) = \dfrac{V_{LP}(0)e^{rt}}{V_{LP}(0)(1/2 + e^{rt}/2)}-1 = \dfrac{2e^{rt}}{1+e^{rt}}-1.
\end{equation*}

\subsection{Required returns for reward LSTs}

Reward-LSTs increase their value every day, according to the staking rate $r$. Therefore, we model their price process as a geometric Brownian motion with drift $r$ and some volatility $\sigma\geq 0$:
\begin{align*}
    P(t) = P(0) \exp\left( \left(r - \frac{\sigma^2}{2}\right)t + \sigma B(t) \right),
\end{align*}
where $\{B(t)\}_{t\geq 0}$ is a standard Brownian motion. A detailed treatment of the price process can be found in Appendix \ref{app:continous_calc}. 
In particular, note that under ideal circumstances where the price of the LST solely reflects the added staking rewards, i.e. the variance $\sigma=0$, the LST's price is given by
\begin{align*}
    P(t) = P(0) e^{rt}.
\end{align*}
In the following, we derive expressions for LP losses under this price process for certain AMMs.

\subsubsection{Required returns for CPMMs.}\label{ssec:retruns_CPMM}

For the reward-LST price process, as described in the previous section, the required returns for trading pairs involving reward-LSTs can be calculated for various AMMs.
For a constant product AMM, we show in Appendix \ref{app:CPMM_returns}, that
\begin{align*}
    rr_{LVH}(t) = \frac{1}{2} \left(e^{(\frac{r}{2} - \frac{\sigma^2}{4})t + \frac{\sigma}{2} B(t)} + e^{(-\frac{r}{2}+\frac{\sigma^2}{4})t - \frac{\sigma}{2}B(t)}\right) -1.
\end{align*}
which in expectation is
\begin{align*}
    \mathbb{E}[rr_{LVH}] 
    = \frac{1}{2}\left(e^{(\frac{r}{2}-\frac{\sigma^2}{8})t} + e^{(-\frac{r}{2}+\frac{3\sigma^2}{8})t}\right)-1.
\end{align*}
For the ideal LST price trajectory with $\sigma=0$, this implies $\mathbb{E}[rr_{LVH}] = (e^{r/2}+e^{-r/2})/2-1$.
For loss-versus-staking, we derive the following expression in Appendix \ref{app:CPMM_returns}:
\begin{align*}
   rr_{LVS}(t) = e^{(\frac{r}{2} - \frac{\sigma^2}{4})t + \frac{\sigma}{2} B(t)} - 1.
\end{align*}
In particular, in expectation, LVS is 
\begin{align*}
    \mathbb{E}\left[rr_{LVS}(t)\right] = e^{(r/2 - \sigma^2/8)t} - 1.
\end{align*}
Finally, for the ideal case, where the LST perfectly tracks the staking rate, i.e.\ $\sigma=0$, we have $\mathbb{E}\left[rr_{LVS}(t)\right] = e^{rt/2} - 1$.

\subsubsection{Required Returns for Concentrated Liquidity CPMMs.}

\begin{figure}[htb]
    \centering
    \includegraphics[width=0.7\textwidth]{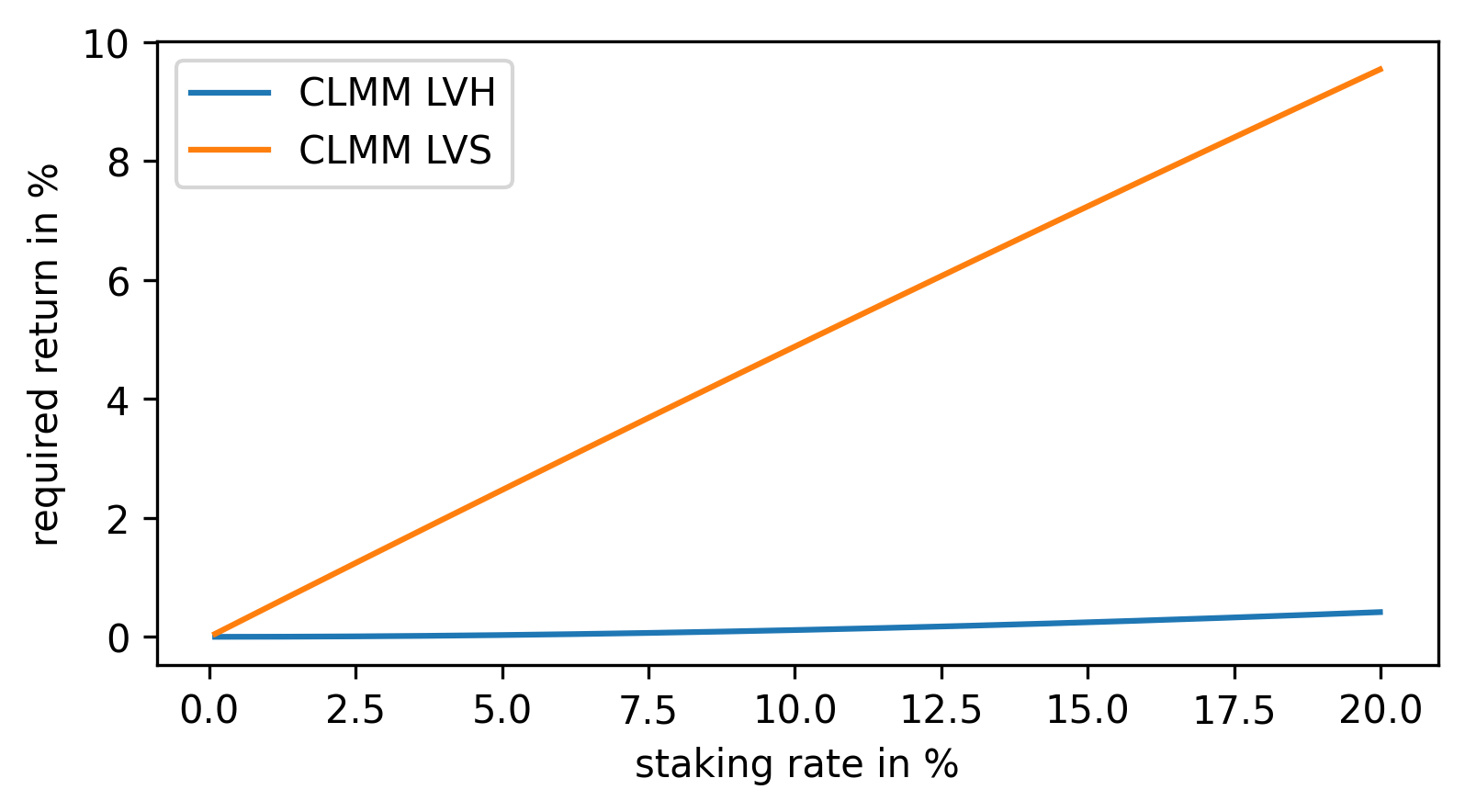}
    \caption{Required returns for different staking rates.}
    \label{fig:required_returns}
\end{figure}

For CLMMs, we consider symmetric positions around the expected price movement of the LST during the time interval $[0,t]$. 
Under the assumption that the LST follows the ideal price trajectory, i.e.\ $P(t) = P(0) e^{rt}$, we derive the closed-form expressions for LVH and LVS in Appendix \ref{app:CLMM_returns}.
For LVH, we find that
\begin{align*}
    rr_{LVH}(t) &= \frac{1}{2}\left( e^{rt/2} + e^{-rt/2} \right) - 1.
\end{align*}
As Figure \ref{fig:required_returns} shows, this value is small for most practical staking rates.
Note also, that this metric is independent of the width of the chosen price range.

Next, we calculate LVS as
\begin{align*}
    rr_{LVS}(t) = \frac{V_{LST}(t)}{V_{LP}(t)} - 1 = e^{rt/2}-1.
\end{align*}
As could be expected, $rr_{LVS}\approx r/2$ (see Figure \ref{fig:required_returns}) stemming from the fact that LPs lose out on the staking rewards on about half of their liquidity.
In other words, returns from fees need to be at least half the staking rate, to make it more attractive to LP in a pool over fully staking one's capital.

Again, a notable discovery is that $rr_{LVS}$ remains unaffected by the chosen width of the position range, provided that the position is symmetric.
This is highly relevant, since it means that the range of the liquidity position can be chosen by solely focussing on maximizing the amount of fees earned.
Achieving this requires optimizing the liquidity concentration versus the duration the position remains within range.

\section{Empirical Analysis}

In this section, we empirically measure the previously described losses to LPs in the largest AMM pools containing liquid staking tokens, and analyze if trading fees have historically compensated LPs adequately.
For each pool, we historically track the value of an LP position that earns fees (i.e. $V_{LP}$ + fees). Moreover, given historical prices, we calculate the values of the HOLD portfolio ($V_{HOLD}$) and a fully staked portfolio ($V_{LST}$).
By comparing $V_{LP}$ with $V_{HOLD}$ and $V_{LST}$, we assess whether LPs earned sufficient compensation to balance LVH and LVS, as described in the previous section.

\subsection{Methodology}

\subsubsection{Data Origin.}
The nine analyzed pools include pools on Uniswap v3 and Curve v1 and v2 on the Ethereum blockchain.

For the Curve pools, historical daily token prices and AMM-pool compositions are sourced from an Ethereum archive node. More specifically, for the last block of each day, we retrieved the token pool reserves via the \emph{balances} method of the Curve pool smart contract and the amount of LP tokens in circulation via the \emph{totalSupply} method of the LP token smart contract. Based on this data, the value of a single LP token for each day and thereby the value of a liquidity position in the pool can be calculated. 
These numbers already include earning from trading fees, since these are paid into the pool. 
On top of trading fees, LPs can earn extra rewards paid in CRV (Curve's protocol token) in certain Curve pools.
We query the value of these rewards from a subgraph using The Graph protocol.\footnote{\url{https://api.thegraph.com/subgraphs/name/messari/curve-finance-ethereum}}
The returns from the CRV rewards are then added to the value of the liquidity position.

For Uniswap v3 pools, we obtain all swap transaction, i.e.\ the amount traded and the amount of active liquidity at the time of the swap, from an Ethereum node. This allows us to historically calculate the fees a liquidity position in the pool earns.

As a reference for the historical staking rate, we use Beacon Chain's \emph{Ether Staking Offered Rate} (ETH.STORE), as it represents the average financial return validators on the Ethereum network have achieved over the last 24 hours. It includes MEV rewards, which are also earned by holders of LSTs \cite{2024ETHStore}.

\subsubsection{Pool Description.}
We analyze nine dominant ETH-LST liquidity pools -- a full list of which can be found in Table \ref{tab:LiquidityPools} in the appendix. Figure \ref{fig:TVLVolume} in the appendix shows that stETH-ETH on Curve v1 has by far the highest TVL among the analyzed pools with about \$250m at the end of 2023, followed by another stETH pool on Curve v1 with about \$70m. The highest daily traded volume on the other hand, is observed for the wstETH-ETH pool on Uniswap v3. The higher trading volume for wstETH in comparison to stETH confirms the higher utilization of wstETH (reward-LST) in DeFi, whereas stETH (reward-LST) tokens seem to be primarily bought to hold and accumulate rewards from staking and consequently less actively traded.

\subsubsection{Computation.}
The daily wealth of LPing ($V_{LP}$ + fees) is calculated using the data described above.
Moreover, the value of HOLD ($V_{HOLD}$) and fully staking ($V_{LST}$) can be computed from the price of the LST.
For each, we assume an initial capital of 1 ETH, and measure the value of the portfolio in ETH.
It is assumed that the staking rewards are re-staked on a daily basis. 

For rebase tokens, such as the stETH pools on Curve v1, we adjust the HOLD and LST portfolios daily to account for the distributions of staking rewards in the form of newly minted LST tokens.

For Uniswap v3 it is assumed that the LP (and HOLD) position is re-balanced on a monthly basis, i.e.\ the price range is reset at the beginning of each month. On each reset, the price range of the liquidity position is set to be $[-0.25\%, +0.75\%]$ around the pool price at the beginning of the month.
This range is based on an expected monthly return from staking of 0.5\% which has approximately been observed on average during our observation period.
Additionally, the range adds $\pm 0.25\%$ around the expected monthly price change to accommodate for some price volatility of the LST.
All studied LSTs pools on Uniswap v3 are pools between reward-LSTs and ETH.

\subsection{Analysis}

\begin{figure*}[!tbp]
  \centering
    \includegraphics[width=0.9\textwidth]{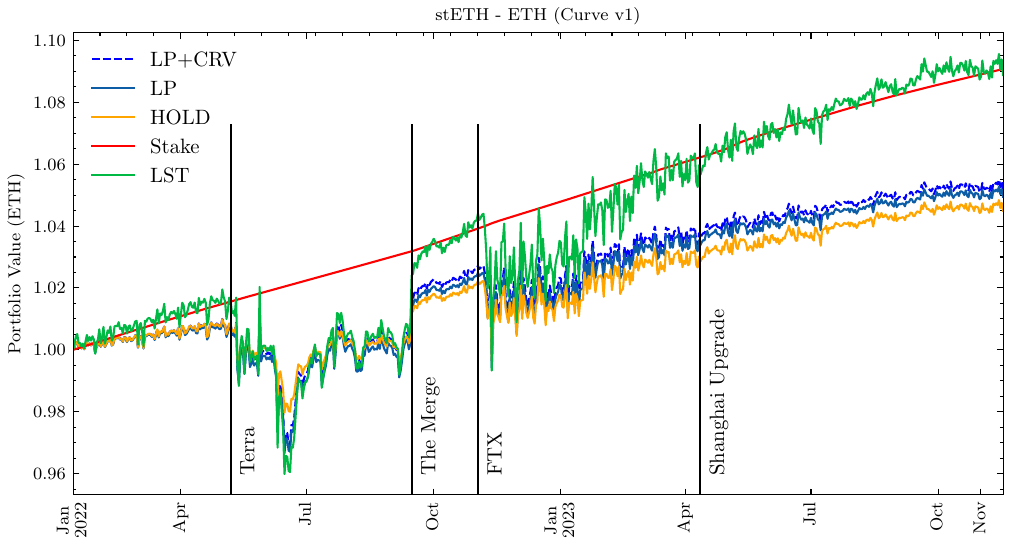}
    \caption{Historical portfolio values for the stETH-ETH pool on Curve v1:  LPing in the pool (blue line), holding the initial position (orange line), and fully holding the LST (blue line). The red line indicates the average staking rate.}
    \label{fig:wealth_stETH_ETH}
\end{figure*}

We begin the analysis with the stETH-ETH pool on Curve v1 - the largest pool in terms of TVL. The development of the considered portfolio values -- LP (including fees), HOLD, and LST (full LST position) -- as well as the reference staking rate are depicted in Figure \ref{fig:wealth_stETH_ETH}. We additionally plot the wealth of an LP including  CRV rewards (LP+CRV). In Curve, LPs earn additional rewards for liquidity provision on top of trading fees. These rewards are distributed in Curve governance tokens - CRV according to its emissions \cite{2024CurveRewards}. While we observe that the additional CRV rewards did not significantly impact the wealth of LPs in this pool, they do make a difference in other pools (see Figure \ref{fig:wealthPools_curve} in the appendix).

We find that during most of the observation period, the value of the liquidity position stays slightly above the value of HOLD. This indicates that fees sufficiently compensate LPs for LVH. On the other hand, the LP's wealth evolves significantly lower than the value of fully staking, meaning that LVS was higher than the returns from fees, and the LPs would have been better off fully investing their portfolio in the LST instead of providing liquidity to the AMM.

\subsubsection{Moving Average.}

\begin{figure*}[!tbp]
  \centering
    \includegraphics[width=0.9\textwidth]{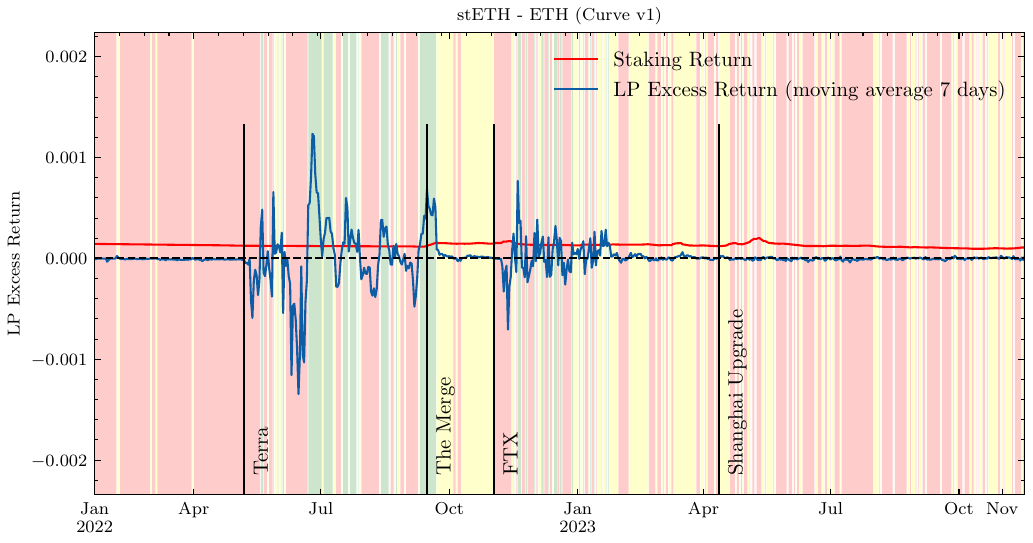}
    \caption{Historical 7 day moving average of LVH and LVS for the stETh-ETH pool on Curve v1. Color green marks periods with no loss-versus-staking (LVS) and color yellow periods with no loss-versus-holding (LVH)}
    \label{fig:MA_stETH_ETH}
\end{figure*}

Whereas Figure \ref{fig:wealth_stETH_ETH} shows the wealth of LP, HOLD, and LST portfolios, assuming they enter the pool at the beginning of the observation period, Figure \ref{fig:MA_stETH_ETH} presents the 7-day moving average of the difference in daily returns of LPs on the one hand, and HOLD and LST on the other hand. The differences indicate whether providing liquidity to the AMMs yielded higher returns than HOLD or LST in the past week. The periods during which LPs' wealth (including fees) grew more than both HOLD and LST are marked in green (fees compensated for LVS). The periods when LP accumulated more wealth than HOLD but less than LST are marked in yellow (fees compensated for LVH, but not for LVS). Periods marked in red denote that LP accumulated less wealth than both HOLD and LST (fees did not compensate for LVH). 

We find that during most of the analyzed periods, liquidity provision resulted in both a loss-versus-holding and a loss-versus-staking over a 7-day period during about half the observation period. Figure \ref{fig:MA30pools_curve} in the appendix, displaying the same comparison, but for a 30-day moving average, shows a similar picture.

\subsubsection{Impact of CRV Rewards.} 
\begin{figure*}[!tbp]
  \centering
    \includegraphics[width=0.9\textwidth]{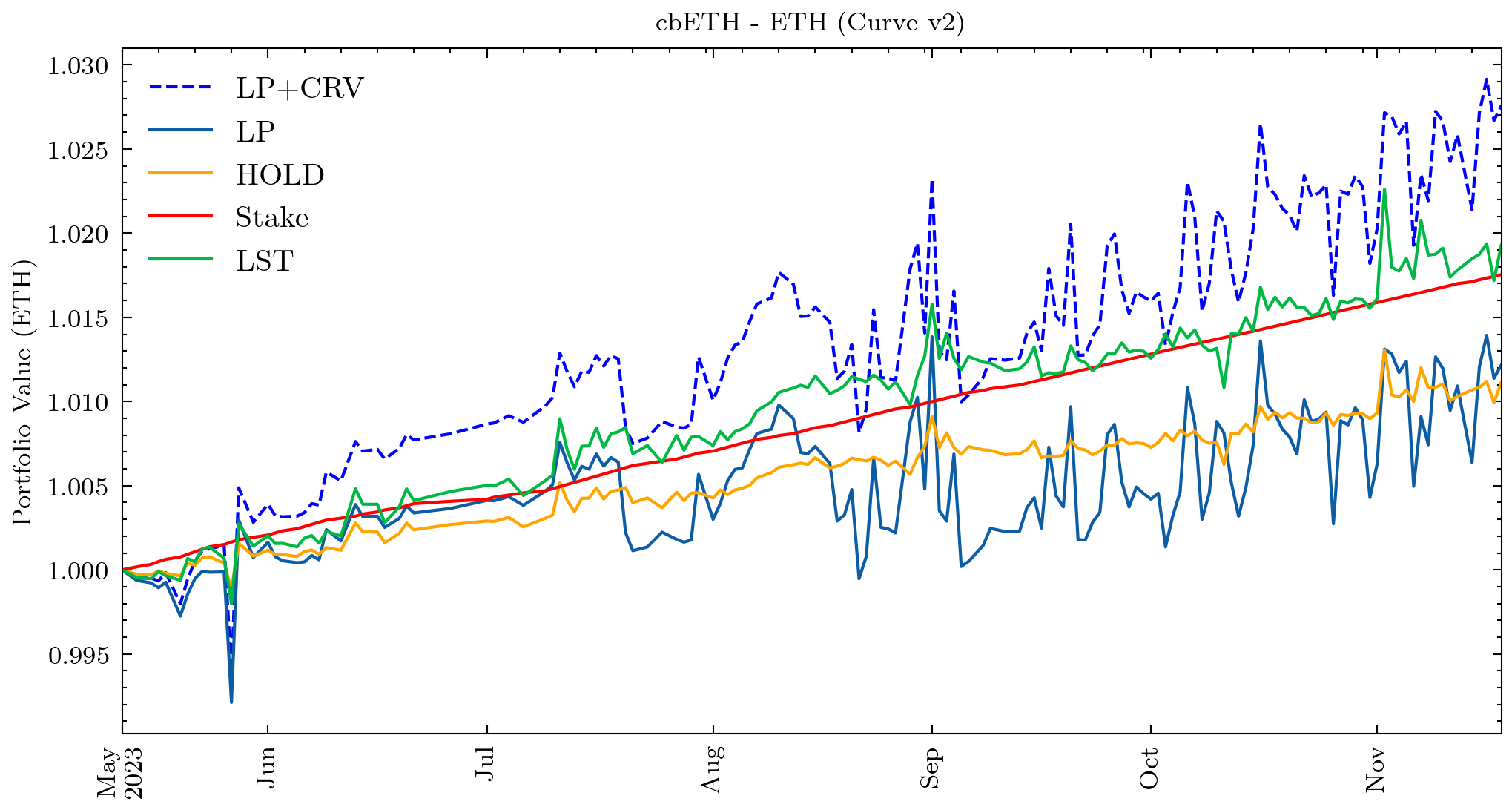}
    \caption{Historical portfolio value of Staker (red line), HOLD (orange line), LPs (blue line) and LST holder (green line) for the cbETH-ETH pool on Curve v2, denominated in ETH}
    \label{fig:wealth_cbETH_ETH}
\end{figure*}

Figure \ref{fig:wealth_cbETH_ETH} exhibits a significantly larger impact of CRV rewards on the profitability of the LP strategy in the cbETH-ETH pool on Curve v2: without CRV rewards, the wealth of LP would not exceed the wealth of HOLD and LST, resulting in loss-versus-holding and loss-versus-staking. With CRV rewards, on the other hand, LPing out earns both the HOLD and LST portfolio.

\subsubsection{LST pools on Curve.}
Figures\ref{fig:wealthPools_curve} and \ref{fig:MA30pools_curve} exhibit the wealth of LP, HOLD, and LST, as well as the 30-day moving averages of their differences, for the five LST pools on Curve. Three of the pools allow swapping stETH for ETH but differ in the parameters used and, consequently, in the behavior of the bonding curve. While the leverage parameter A for the largest pool in terms of TVL is 30, the ``concentrated'' (A=1000) and ``ng'' (A=1500) pools that have later inception dates, offer more concentrated liquidity. Figure \ref{fig:MA30pools_curve} shows that the more concentrated pools almost always compensate LPs for loss-versus-holding, but trading fees are not sufficient to compensate for loss-versus-staking.

The analysis also includes two pools with reward-LSTs -- rETH and cbETH -- deployed to Curve v2. As reward-LSTs, rETH and cbETH are not pegged to 1 ETH and their pools use the Cryptoswap Invariant - an AMM that concentrates liquidity around their current prices.
The returns in the rETH and cbETH pools were significantly impacted by the distribution of CRV token rewards. As depicted in Figure \ref{fig:wealthPools_curve} (blue dotted lines), the LP wealth including CRV tokens actually significantly exceeded the staking returns for rETH and cbETH.
Without rewards, the picture is similar to stETH pools, where trading fees are not sufficient to compensate for loss-versus-staking.

\subsubsection{LST pools on Uniswap.}
Figures \ref{fig:wealthPools_uniswap}, and \ref{fig:MA30pools_uniswap} show the wealth of LP, HOLD, and LST, as well as the 30-day moving average of wealth difference, for the four LST pools on Uniswap v3. Due to incompatibility with rebase tokens, all LSTs listed at Uniswap are reward-based (wstETH, rETH, and cbETH). 
The moving averages plotted in Figure \ref{fig:MA30pools_uniswap} shows that the pools with the higher fee rate of 5 basis points were able to generate a sufficient amount of fees to compensate for LVS during certain time periods, while the 1 basis point wstETH-ETH pool consistently failed to achieve this.

Over the whole observation period, total fee returns were generally high enough to compensate LPs for LVS in the 5 basis point pools, especially in the more recent past (see right plots in Figure \ref{fig:wealthPools_uniswap}).
However, the actual profitability of liquidity positions depends on their holding period, as it is affected by price fluctuations of cbETH and rETH.  
In particular, depositing into the pool during the depegs following the FTX insolvency in November 2022 (rETH being overpriced while cbETH was underpriced), can lead to different results.

\subsection{Discussion}

The primary objective of the empirical part was to assess the profitability of allocating LSTs to AMMS.
We found that the majority of LST liquidity is currently not sufficiently compensated compared to the possibility of fully staking their capital.
This puts the sustainability of the current allocation of LST liquidity on AMMs into question.
If return from trading fees do not increase, e.g.\ as a result of increased trading volume, the current amount of liquidity will likely not remain in these pools in the long term.

\subsubsection{CRV rewards.}
We find that most Curve pools currently are only profitable for LPs, especially compared to fully staking, when taking the CRV token rewards to LPs into account. The amounts of CRV token rewards emitted to pools are decided by the Curve governance, and can change every week. A drop in their distribution to the LST pools, could lead to lower returns for LPs and subsequently a decrease in liquidity in these pools. 

\subsubsection{CLMM Rebalancing.}
LP returns on CLMMs such as Uniswap v3 can be increased by choosing a smaller price range to provide liquidity to and rebalance more frequently.
However, this strategy requires active monitoring of the LP position and incurs gas fees, thereby reducing LP's profits. Additionally, we observed the emergence of new pools featuring the same LST and ETH pairs but offering lower trading fees (1 basis point instead of 5 basis points). Despite attracting higher trading volumes, these new pools so far do not generate a sufficient compensation for LPs.

\subsubsection{Reward-LST token pairs.}
One possible approach to reduce LP losses is to utilize pools involving two reward-LSTs, such as wstETH and rETH.
Such pools would reduce LVH, since the LST prices should move in parallel with staking rewards, and LVS, since the full liquidity position earns trading rewards.
Moreover, one reward-LST in such pools could potentially profit from the deep liquidity of the other reward-LST to the underlying.
However, at the time of writing, these pools have so far not attracted much liquidity or trading volume.

\subsubsection{Limitation of Rebase-LST.}
It is notable that stETH, the 7th largest cryptocurrency by marketcap at the time of writing \cite{2023CoinGecko}, is not traded on Uniswap. As a rebase-LST, stETH's price is pegged to 1 ETH, meaning an stETH-ETH AMM pool would potentially not suffer impermanent loss. However, the current implementation of Uniswap protocol does not support pools with rebase tokens, and listing of stETH is not possible.

\section{Conclusion} 

Liquid staking tokens as liquidity on AMMs has grown to a vital part of the DeFi landscape, but has so far seen limited academic attention.
This work systemizes liquidity provision of LSTs on AMMs, and both theoretically and empirically studies its returns.

In addition to LVH (``impermanent loss''), the most commonly used metrics to measure LP losses on AMM, we introduce loss-versus-staking (LVS) which is specifically tailored to the case of LST pools on AMMs.
Our empirical results indicate that most AMM pools with LSTs have historically not compensated for loss-versus-staking. Nevertheless, most pools did compensate for LVH (``impermanent loss'').

This could lead to less LST liquidity being provided to AMMs in the long term.
Currently, LST liquidity is already low compared to the overall LST supply (\$0.5 billion vs. over \$30 billion). A further decrease could negatively impact the price stability of LSTs and increase their risk of depegging.

\section*{Acknowledgements} 
This research article is a work of scholarship and reflects the authors' own views and opinions. It does not necessarily reflect the views or opinions of any other person or organization, including the authors' employer. Readers should not rely on this article for making strategic or commercial decisions, and the authors are not responsible for any losses that may result from such use. 
One of the authors' research was supported by the \textit{EPSRC CDT in Mathematics of Random Systems} (EPSRC Grant  EP/S023925/1).

\bibliographystyle{splncs04}
\bibliography{main}

\appendix

\newpage
\section{Continuous-time Loss-versus-staking}\label{app:continous_calc}

Assume there exists an infinitely deep centralized exchange, where the token $T$ and the LST can be traded without any fees. The price of the token is given by a geometric Brownian motion  that is a $\mathbb{Q}$-martingale, i.e.\
$$\frac{d P_T(t)}{P_T(t)} = \mu_T(t) dt + \sigma_T(t) dW_T \qquad \forall t\geq 0,$$ with $\mu_T(t)$ being the long-term drift of the token and $\sigma_T$ the volatility of the Token. For the LST we have 
$$\frac{d P_{LST}(t)}{P_{LST}(t)} = \mu_{LST}(t) dt + \sigma_{LST}(t) dW_{LST} \qquad\forall t\geq 0.$$
Assuming the LST works as intended on the long run, its drift $\mu^{LST}(t)$ would be composed of $T$'s drift $\mu_{T}(t)$ plus the staking rate $r$, i.e.\ $\mu_{LST}(t) = \mu_{T}(t) + r$. In addition, the two tokens are clearly not independent and have some sort of correlation $\rho$, resulting in two correlated Brownian motions $W_T$ and $W_{LST}$ with quadratic variation $[W_T, W_{LST}](t) = \rho(t)$.

\noindent Then we can rewrite the two Brownian motions as
\begin{align*}
    W_T &= B_T \\
    W_{LST} &= \rho B_T + \sqrt{1-\rho^2}B_{LST}
\end{align*}
and get the analytic solution (It$\hat{o}$'s formula on $ln(P_{T/LST}(t))$): 
\begin{align*}
    P_T(t) &= P_T(0)\exp((\mu_T - \frac{1}{2}\sigma_1^2)t+\sigma_1 B_T(t)) \\
    P_{LST}(t) &= P_{LST}(0)\exp((\mu_{LST} - \frac{1}{2}\sigma_1^2)t+\sigma_2(\rho B_{T}(t) +  \sqrt{1-\rho^2}B_{LST}(t)))
\end{align*}

In our paper we focus only on pairs of LST and ETH as the underlying token T. So we don't compare it to a third currency and therefore can omit the GBM from the token, resulting in the much simpler stochastic process for LST: 
$$\frac{d P_{LST}(t)}{P_{LST}(t)} = \underbrace{\mu_{LST/T}}_{=r} dt + \sigma_{LST/T}(t) dB(t) \qquad \forall t\geq 0,$$
where $\mu_{LST/T}$ is the drift of the LST against the token, which is given by the staking rate $r$, $\sigma_{LST/T}$ is the historical standard deviation of the price of LST in T and $B(t)$ a standard Brownian motion.\\

\noindent We then can define a self-financing trading strategy for our staking portfolio with LST that starts holding only LST (1,0) at the beginning and keeps holding them by
$$V_{LST}(t) = V(0)+ \int\limits_0^t  1 dP_{LST} \qquad \forall t \geq 0.$$
Since the portfolio only consists of LST this can be simplified to the initial value times the change in price:
$$V_{LST}(t) = V(0) \exp((r - \frac{1}{2}\sigma_{LST/T}^2)t+\sigma_{LST/T} B(t))$$
leading to an expected value of 
$$
\mathbb{E}[V_{LST}(t)] = V(0)  \exp((r - \frac{1}{2}\sigma_{LST/T}^2)t)+\frac{1}{2}\sigma_{LST/T}^2t = V_0 \exp(rt),
$$
where we used that $\mathbb{E}[\exp(uZ)] = \exp(\frac{u^2}{2})$ for $u\in\mathbb{R}$ and $Z\sim \mathcal{N}(0,1)$.
Hence, the LVS is given by
\begin{align*}
LVS(t) = V_{LST}(t) - V_{LP}(t) = V(0)\exp((r - \frac{1}{2}\sigma_{LST/T}^2)t+\sigma_{LST/T} B(t)) - V(t).
\end{align*}
Here, the value of $V_{LP}(t)$ depends on the chosen AMM and corresponding parameters. We discuss two examples in the following sections.

\section{LP Losses in CPMMs}\label{app:CPMM_returns}

For the standard CPMM with 2 assets, as described in \cite{Adams2020UniswapCore}, the holdings of a liquidity position are
\begin{align*}
    x(t) = \frac{L}{\sqrt{P(t)}} \qquad \text{and}\qquad y(t) = L\sqrt{P(t)},
\end{align*}
where $L$ is the size of the position.
Hence, we have the following portfolio values:
\begin{align*}
    V_{LP}(t) &= x(t)P(t) + y(t) = 2L\sqrt{P(t)} \\
    V_{HOLD}(t) &= x(0)P(t) + y(0) = \frac{L}{\sqrt{P(0)}}P(t) + L\sqrt{P(0)} \\
    V_{LST}(t) &= \left(x(0) + \frac{y(0)}{P(0)} \right) P(t) = \frac{2L}{\sqrt{P(0)}} P(t)
\end{align*}

\subsubsection{Loss-versus-staking.}
The required return to compensate LVS is
\begin{align*}
    rr_{LVS}(t) = \frac{V_{LVS}(t)}{V_{LP}(t)} - 1  = \frac{\sqrt{P(t)}}{\sqrt{P(0)}} - 1.
\end{align*}
To derive $\sqrt{P(t)}$, we apply Ito's formula with $f(t,P_{LST}(t)) = \sqrt{P_{LST}(t)}$:
\begin{align*}
    d\sqrt{P(t)} &= \frac{1}{2\sqrt{P(t)}}dP(t) + \frac{1}{2}\cdot\frac{-1}{4}P(t)^{\frac{-3}{2}}(dP(t))^2\\
    &= \frac{1}{2\sqrt{P(t)}}(r P(t) dt + \sigma P(t) dB(t)) - \frac{1}{8}P(t)^{\frac{-3}{2}} (\sigma^2\cdot P(t)^2) dt \\
    &= (\frac{r}{2} \sqrt{P(t)} - \frac{\sigma^2}{8}\sqrt{P(t)}) dt + \frac{\sigma}{2} \sqrt{P(t)} dB(t))
\end{align*}
This implies
\begin{align*}
    \sqrt{P(t)} &= \sqrt{P(0)}\cdot \exp((\frac{r}{2} - \frac{\sigma^2}{8} - \frac{1}{2}(\frac{\sigma}{2})^2)t + \frac{\sigma}{2} B(t)) \\
    &= \sqrt{P(0)}\cdot \exp((\frac{r}{2} - \frac{\sigma^2}{4})t + \frac{\sigma}{2} B(t)).
\end{align*}
Inserting this into loss-versus-staking, yields
\begin{align*}
    rr_{LVS}(t) = e^{(\frac{r}{2} - \frac{\sigma^2}{4})t + \frac{\sigma}{2} B(t)} - 1.
\end{align*}
In particular, in expectation, LVS is 
\begin{align*}
    \mathbb{E}\left[rr_{LVS}(t)\right] = e^{(\frac{r}{2} -\frac{\sigma^2}{4})t} \mathbb{E}[\exp(\frac{\sigma}{2} B(t))] - 1
    = e^{(\frac{r}{2} -\frac{\sigma^2}{8})t} - 1.
\end{align*}
Finally, note that in the ideal case, where the LST perfectly tracks the staking rate, i.e.\ $\sigma=0$, we have $\mathbb{E}\left[rr_{LVS}(t)\right] = e^{rt/2} - 1$.

\subsubsection{Loss-versus-holding.}
The required return to compensate LVH is
\begin{align*}
    rr_{LVH} = \frac{1}{2}\left(\frac{\sqrt{P(t)}}{\sqrt{P(0)}} + \frac{\sqrt{P(0)}}{\sqrt{P(t)}}\right) - 1
\end{align*}
To calculate the change in LVH we need the differential $\frac{1}{\sqrt{P(t)}}.$ We use Ito's formula with $f(t,P(t)) = \frac{1}{\sqrt{P(t)}}$ and get:
\begin{align*}
    d\frac{1}{\sqrt{P(t)}} & = (-\frac{1}{2}P(t)^{-\frac{3}{2}} dP(t) + \frac{1}{2}\cdot\frac{3}{4}P(t)^{-\frac{5}{2}}) (dP(t))^2\\
    &= (-\frac{1}{2}P(t)^{-\frac{3}{2}} (r P(t) dt + \sigma P(t) dB(t))  + \frac{3}{8}P(t)^{-\frac{5}{2}} (\sigma^2 P(t)^2) dt  \\  
    &= (\frac{3\sigma^2}{8\sqrt{P(t)}}-\frac{r}{2\sqrt{P(t)}})dt - \frac{\sigma}{2\sqrt{P(t)}} d B(t)
\end{align*}
which implies: 
\begin{align*}
    \frac{1}{\sqrt{P(t)}} &= \frac{1}{\sqrt{P(0)}}\cdot\exp((-\frac{r}{2}+\frac{3\sigma^2}{8}-\frac{\sigma^2}{8})t - \frac{\sigma}{2}B(t))) \\
    &=\frac{1}{\sqrt{P(0)}}\cdot\exp((-\frac{r}{2}+\frac{\sigma^2}{4})t - \frac{\sigma}{2}B(t)))
\end{align*}
This leads to the following expression for the required return:
\begin{align*}
    rr_{LVH} = \frac{1}{2} \left(e^{(\frac{r}{2} - \frac{\sigma^2}{4})t + \frac{\sigma}{2} B(t)} + e^{(-\frac{r}{2}+\frac{\sigma^2}{4})t - \frac{\sigma}{2}B(t)}\right) -1
\end{align*}
which in expectation is
\begin{align*}
    \mathbb{E}[rr_{LVH}] 
    = \frac{1}{2}\left(e^{(\frac{r}{2}-\frac{\sigma^2}{8})t} + e^{(-\frac{r}{2}+\frac{3\sigma^2}{8})t}\right)-1.
\end{align*}
And for $\sigma=0$, we have $\mathbb{E}[rr_{LVH}] = (e^{rt/2}+e^{-rt/2})/2-1$.

\section{LP Losses in Concentrated Liquidity CPMMs}\label{app:CLMM_returns}

In the following, we examine LP's losses and required returns for CPMM with concentrated liquidity as introduced in \cite{Adams2021UniswapCore}.

Consider a concentrated liquidity position that is opened at time $t=0$ with the prospect of maintaining it until time $t=T$.
The range of the positions is set symmetric around the expected price movement from $P(0)$ to $P(0)e^{rT}$ during that period, i.e.\ as $[P(0)e^{-d}, P(0)e^{rT+d}]$ for some $d\geq 0$.
Hence, if the price is range at time $t\geq 0$, the position has the following holdings, where $L$ denotes the size of the liquidity position.
\begin{align*}
    x(t) &= L\left( \frac{1}{\sqrt{P(t)}} - \frac{1}{\sqrt{P(0)e^{rT+d}}} \right)\\
    y(t) &= L\left( \sqrt{P(t)} - \sqrt{P(0)e^{-d}} \right)
\end{align*}
In particular, if the LST functions as intended and its price has increased according to the staking rate at the end of the period, such that $P(T) = P(0)e^{rT}$, we have
\begin{align*}
    V_{LP}(T) = x(T)P(T) + y(T) = L\sqrt{P(0)}\left( 2e^{rT/2} - e^{rT/2-d/2} - e^{-d/2} \right).
\end{align*}

\subsubsection{Loss-versus-staking.}
On the other hand, holding only the LST results in a final wealth of
\begin{align*}
    V_{LST}(T) = \left( x(0) + \frac{y(0)}{P(0)} \right) P(T)
    = L\sqrt{P(0)}\left(2 - e^{-rT/2-d/2} - e^{-d/2}\right)e^rT.
\end{align*}
This implies
\begin{align*}
    rr_{LVS}(T) = \frac{V_{LST}(T)}{V_{LP}(T)} - 1 = e^{rT/2}-1.
\end{align*}

In particular, note that $rr_{LVS}$ is independent of the position width $d$, and depends only on $r$ (under the assumption that the price of LST increases according to the staking rate).

\subsubsection{Loss-versus-holding.}

Note that the initial position was chosen to take the expected price change of the LST into account. It was set symmetric around the expected price, and not symmetric around the initial price, as would have been the default without an expected price movement.
With the default range symmetric around the initial price, both the initial position and the expected holding of the liquidity position over a future time period will be split 50-50 in value between the two tokens in the pool.
In particular, considering LVH means comparing the liquidity position to a portfolio with the same expected holdings.

To achieve the same for our setting with an expected price change and an accordingly adjusted initial position, we also adjust the HOLD portfolio:
We consider the position $(\overline{x}(0), \overline{y}(0))$ whose initial holdings equal the expected holdings of the liquidity position $(\overline{x}(P(0)), \overline{y}(P(0))$. 
Let $\overline{V_{HOLD}}$ denote the value of this position.
Specifically, for a symmetric position around the expected movement of the LST, the average (expected) holdings of the liquidity position over the time period is 50-50.
Hence, for adjusted LVH, we compare to holding a position that starts with 50\% value in each ETH and the LST. Simple calculation show that the initial composition of such a 50-50 position (that has the same value as the liquidity position at $t=0$) is
\begin{align*}
    \overline{x}(0) = \frac{L}{\sqrt{P(0)}} \frac{1}{2} \left(2 - e^{-rT/2-d/2} - e^{-d/2} \right),\\
    \overline{y}(0) = L \sqrt{P(0)} \frac{1}{2} \left(2 - e^{-rT/2-d/2} - e^{-d/2} \right).
\end{align*}
From this, the wealth of holding
\begin{align*}
    \overline{V_{HOLD}}(T) = \overline{x}(0) P(T) + \overline{y}(0)
    = L\sqrt{P(0)} \frac{1}{2} \left(2 - e^{-rT/2-d/2} - e^{-d/2} \right) \left( 1+e^{rT} \right)
\end{align*}
and, in turn, the loss-versus-holding can be calculated:
\begin{align*}
    rr_{LVH}(T) &= \frac{\overline{V_{HOLD}}(T)}{V_{LP}(T)} - 1 = \frac{1}{2}\left( e^{rT/2} + e^{-rT/2} \right) - 1
\end{align*}
Again, note that $rr_{LVH}$ simplifies to be independent of $d$, and depends solely on $r$.

\section{Extra Figures and Tables}

\begin{figure*}[!h]
  \centering
    \includegraphics[width=0.9\textwidth]{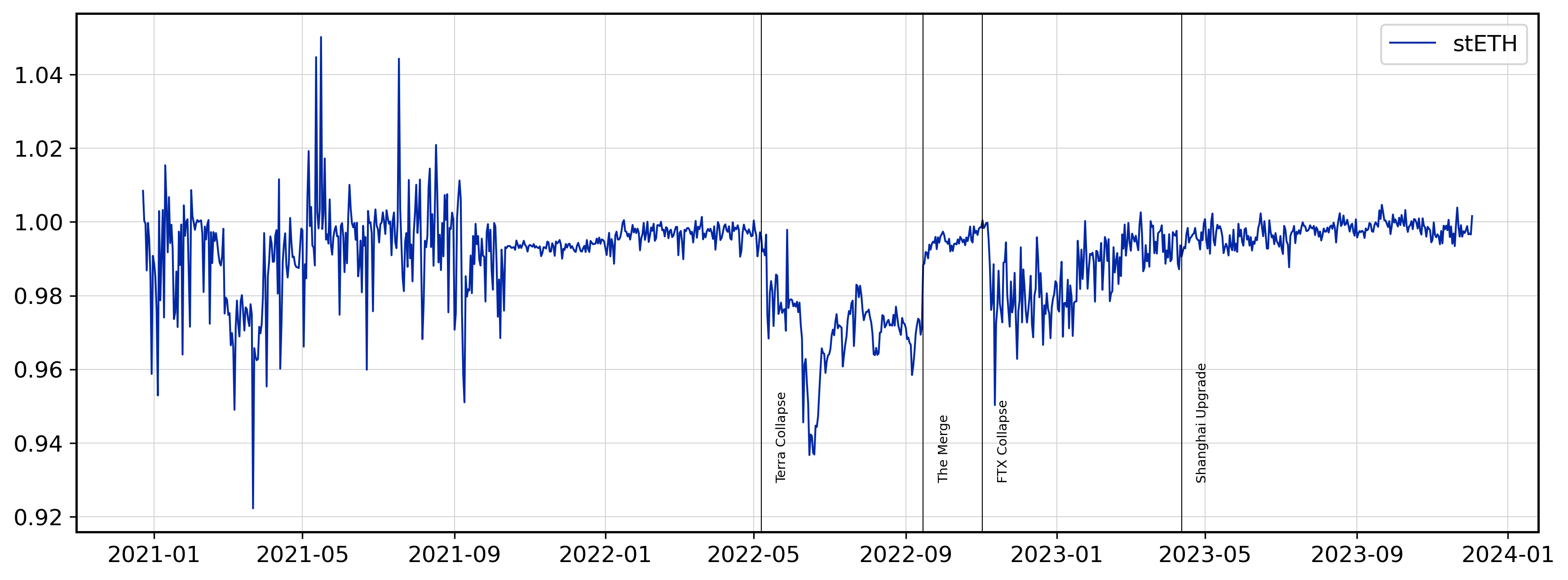}
    \caption{Historical market value of stETH - rebase-LSTs pegged to 1 ETH, denominated in ETH}
    \label{fig:RebaseLST}
\end{figure*}

\begin{figure*}[!h]
  \centering
    \includegraphics[width=0.9\textwidth]{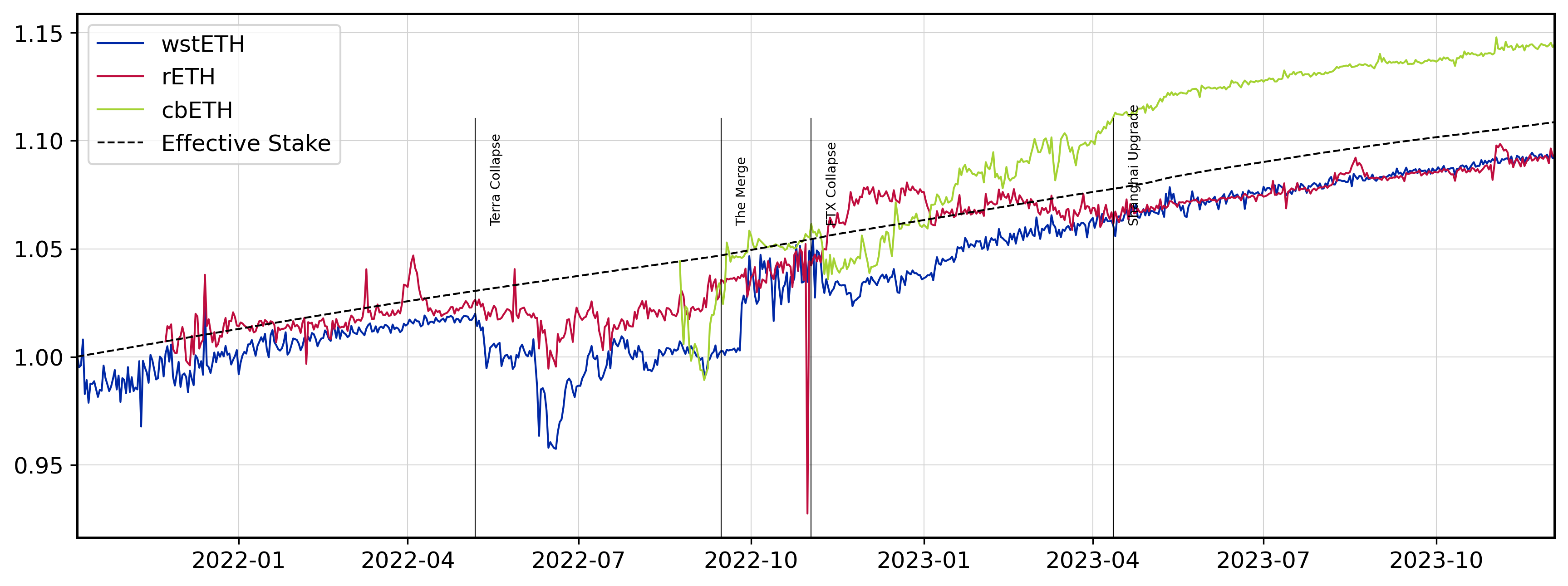}
    \caption{Historical daily returns of reward LSTs, compared to daily returns from staking ETH (dashed line)}
    \label{fig:RewardsLST}
\end{figure*}

\begin{table*}[!bp]
\centering
\caption{Overview of liquidity pools analyzed in this work. Total value locked (TVL) and 24h trading volume denominated in USD on 31st December 2023. The parameters A and Gamma are constants in Stableswap and Cryptoswap Invariant AMMs}
\begin{tabularx}{\textwidth}{lXlXXl}
\textbf{Pool}        & \textbf{AMM} & \textbf{A}  & \textbf{Gamma} & \textbf{TVL} & \textbf{24h Volume} \\
\toprule 
wstETH\_WETH\_100       & Uniswap v3                   &     -        &     -       & 39.66m       & 58.65m              \\
wstETH\_WETH\_500       & Uniswap v3                    &    -         &    -        & 0.77m        & 0.02m                   \\
cbETH\_WETH\_500       & Uniswap v3                  &      -       &      -      & 4.01m       & 0.7m               \\
rETH\_WETH\_500        & Uniswap v3                  &     -        &    -        & 4.53m        & 0.3m               \\
stETH\_WETH             & Curve v1                & 30          &       -     & 251.64m      & 17.72m              \\
stETH\_WETH\_ng        & Curve v1                 & 1500        &   -         & 74.61m       & 1.25m               \\
stETH\_WETH\_con        & Curve v1                 & 1000        &   -         & 4.35m       & 0.12m               \\
rETH\_WETH             & Curve v2                & 20 000 000  & 0.01       & 8.18m       & 0.6m               \\
cbETH\_WETH            & Curve v2                 & 20 000 000  & 0.01       & 4.29m        & 0.30m               \\
\bottomrule
\end{tabularx}
\label{tab:LiquidityPools}
\end{table*}

\begin{figure*}[!tbp]
  \centering
  \makebox[\textwidth][c]{\includegraphics[width=\textwidth]{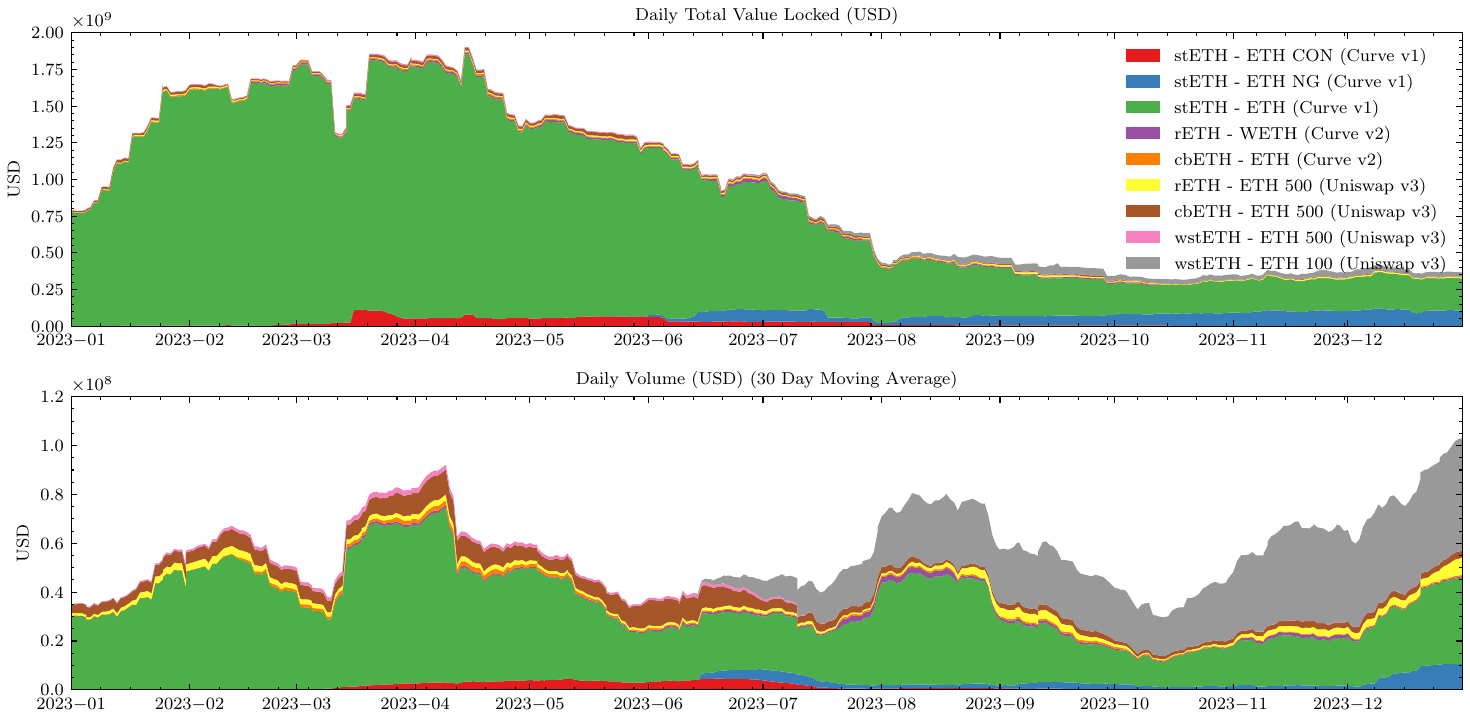}}
    \caption{Hisotrical daily TVL and daily volume (30 day moving average) for analyzed liquidity pools with LSTs, denominated in ETH}
    \label{fig:TVLVolume}
\end{figure*}

\begin{figure*}[!tbp]
  \centering
  \makebox[\textwidth][c]{%
  \begin{subfigure}{1.11\textwidth}
    \includegraphics[width=\textwidth]{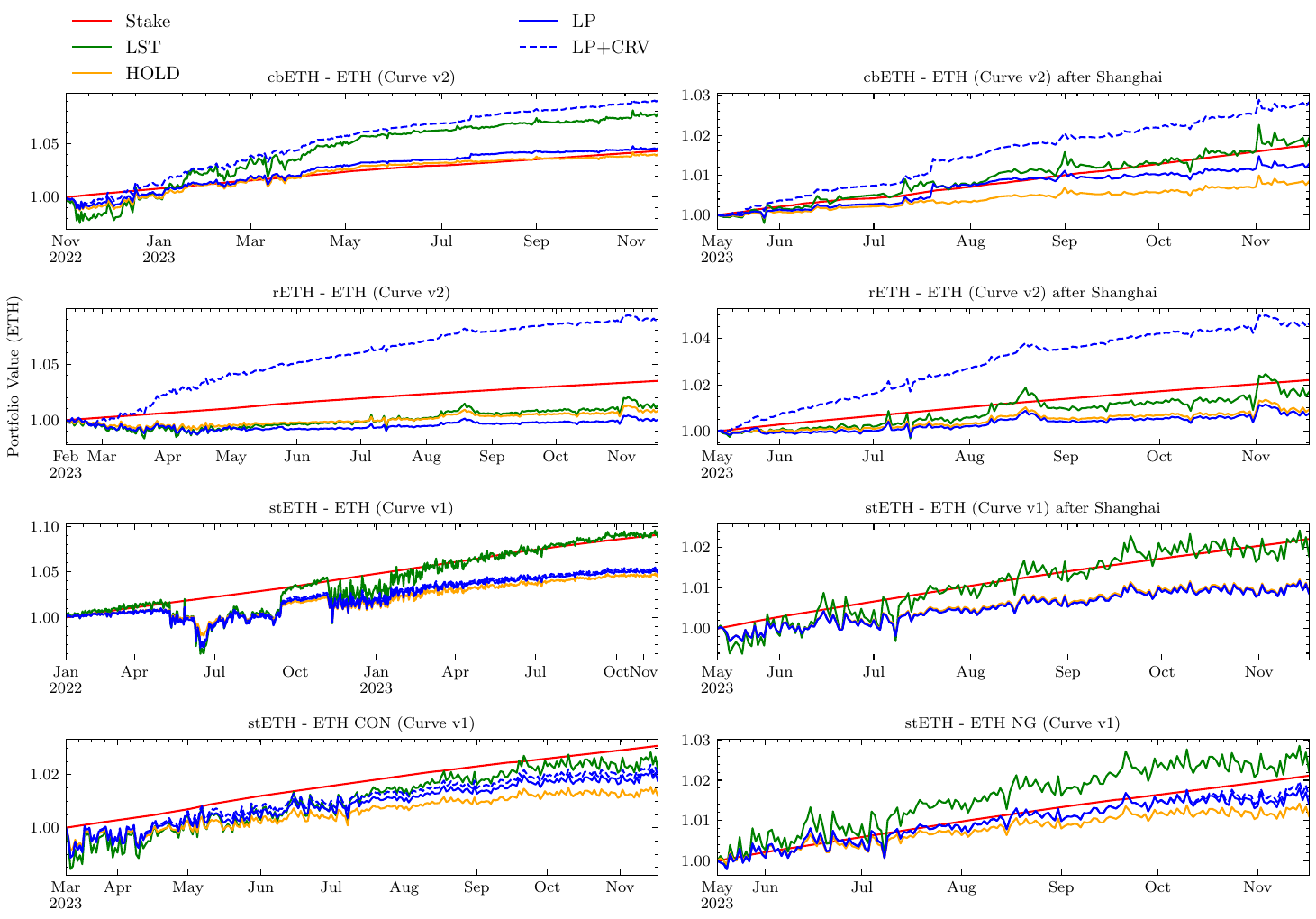}
    \caption{Curve pools}
    \label{fig:wealthPools_curve}
  \end{subfigure}%
  }
  \makebox[\textwidth][c]{%
  \begin{subfigure}{1.11\textwidth}
  \centering
    \includegraphics[width=\textwidth]{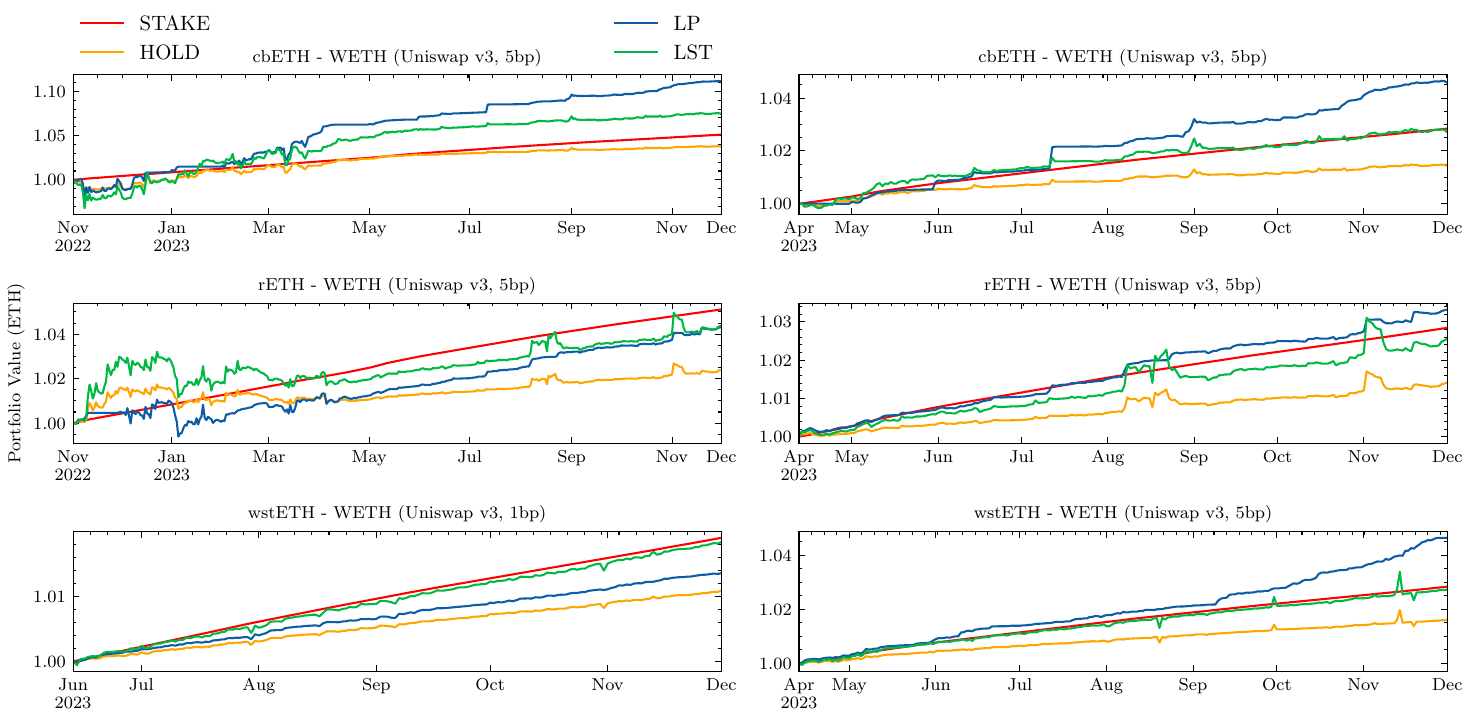}
    \caption{Uniswap v3 pools}
    \label{fig:wealthPools_uniswap}
  \end{subfigure}%
  }
  \caption{Historical portfolio value of LPs (blue line), LPs with CRV token rewards (dotted blue line), HOLD (orange line), LST holder (green line), and the average staker (red line) for selected Curve and Uniswap v3 LST pools, denominated in ETH. Pools with a long history are additionally plotted after the Shanghai upgrade.}
\end{figure*}

\begin{figure*}[!tbp]
  \centering
  \makebox[\textwidth][c]{%
  \begin{subfigure}{1.11\textwidth}
    \centering
    \includegraphics[width=\textwidth]{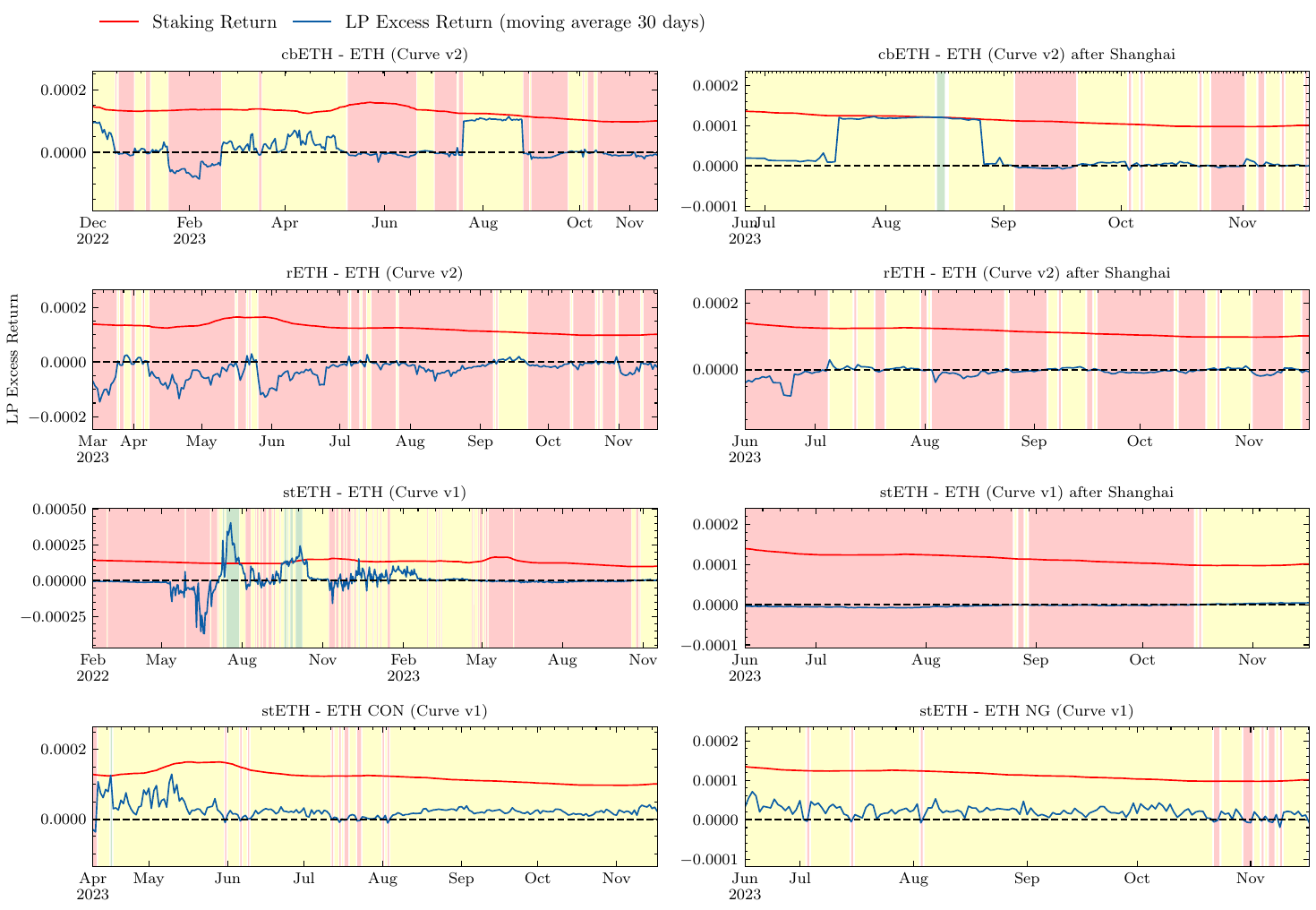}
    \caption{Curve pools}
    \label{fig:MA30pools_curve}
  \end{subfigure}%
  }
  \makebox[\textwidth][c]{%
  \begin{subfigure}{1.11\textwidth}
    \centering
    \includegraphics[width=\textwidth]{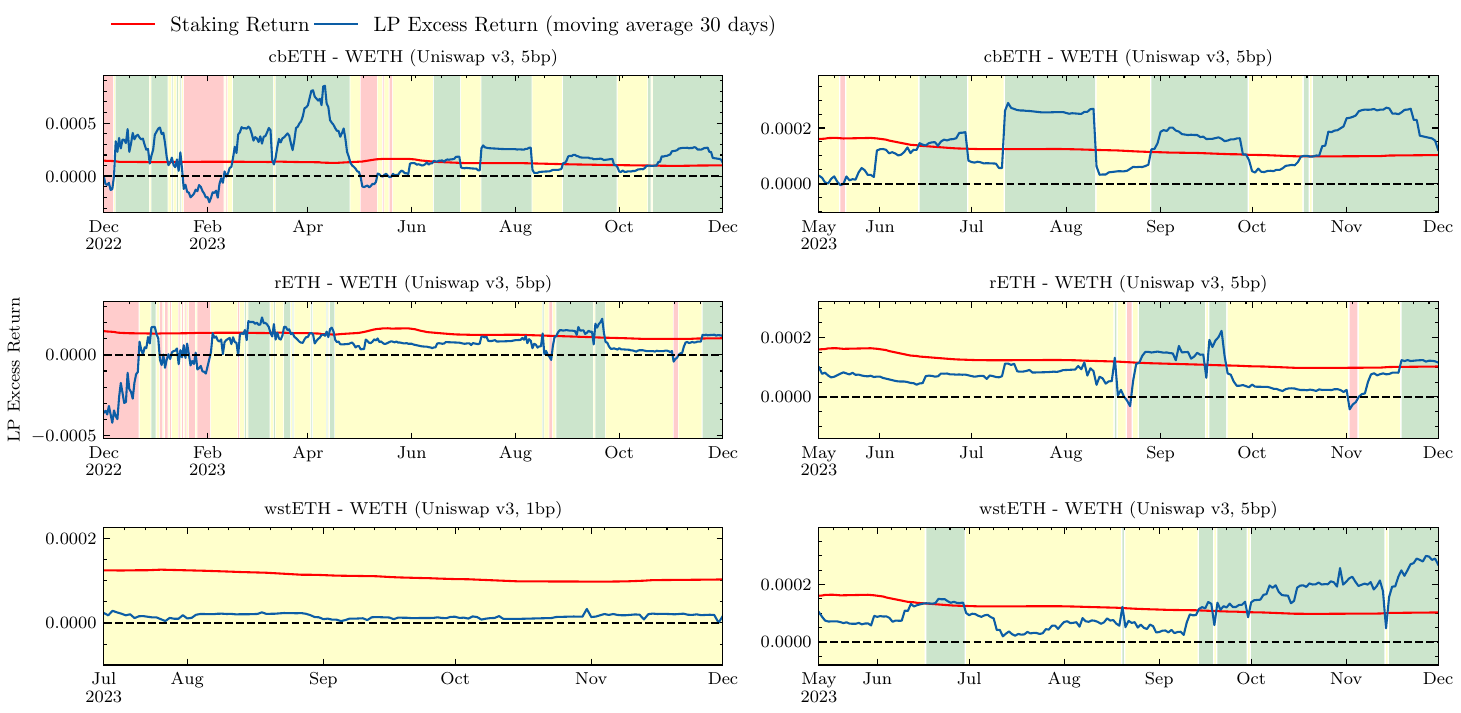}
    \caption{Uniswap v3 pools}
    \label{fig:MA30pools_uniswap}
  \end{subfigure}%
  }
  \caption{Historical 30 day moving average for selected Curve and Uniswap v3 LST liquidity pools. Color green marks periods with no loss-versus-staking (LVS) and color yellow periods with no loss-versus-holding (LVH, impermanent loss). Pools with a longer history are additionally plotted after the Shanghai upgrade.}
\end{figure*}

\end{document}